# Agglomeration Drives The Reversed Fractionation of Aqueous Carbonate and Bicarbonate at the Air-Water Interface

Shane W. Devlin[†,‡], Amanda A. Chen[§], Sasawat Jamnuch[§], Qiang Xu[‡], , Jin Qian[‡], Tod A. Pascal[*,§,1,2], Richard J. Saykally[* †,‡]

†: Department of Chemistry, University of California, Berkeley, CA 94720.
‡: Chemical Sciences Division, Lawrence Berkeley National Lab, Berkeley, CA 94720.

§: ATLAS Materials Science Laboratory, Department of Nano Engineering and Chemical Engineering, University of California, San Diego, La Jolla, California, 92023, USA

1. Materials Science and Engineering, University of California San Diego, La Jolla, California, 92023, USA
2. Sustainable Power and Energy Center, University of California San Diego, La Jolla, California, 92023, USA

*: Corresponding Author: Richard J. Saykally: saykally@berkeley.edu
Tod A. Pascal tpascal@ucsd.edu

**Abstract**: In the course of our investigations of the adsorption of ions to the air-water interface, we previously reported the surprising result that doubly-charged carbonate anions exhibit a stronger surface affinity than do singly-charged bicarbonate anions. In contrast to monovalent, weakly hydrated anions, which generally show enhanced concentrations in the interfacial region, multivalent (and strongly hydrated) anions are expected to show much weaker surface propensity. In the present work, we use resonantly enhanced deep-UV second harmonic generation spectroscopy to measure the Gibbs free energy of adsorption of both carbonate ($CO_3^{2-}$) and bicarbonate ($HCO_3^-$) anions to the air-water interface. Contrasting the predictions of classical electrostatic theory, and in support of our previous findings from X-ray photoelectron spectroscopy, we find that carbonate anions do indeed exhibit much stronger surface affinity than do the bicarbonate anions. Molecular dynamics simulation reveals that strong ion pairing of $CO_3^{2-}$ with the $Na^+$ counter-cation in the interfacial region, resulting in formation of near-neutral agglomerates of $Na^+$ and $CO_3^{2-}$ clusters, is responsible for this counterintuitive behavior. These findings not only advance our fundamental understanding of ion adsorption chemistry, but will also impact important practical processes such as ocean acidification, sea-spray aerosol chemistry, and mammalian respiration physiology.

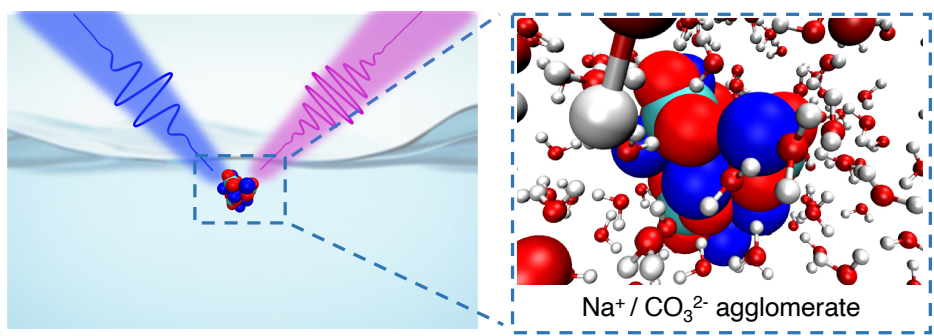

Na⁺ / CO₃²⁻ agglomerate

Introduction

Chemistry occurring at aqueous interfaces governs many important phenomena, e.g. reactions in atmospheric aerosols[1,2] and the uptake of gases at the ocean surface[3], as well as myriad biological processes. Reactions occurring at such an interface are often quite distinct from the same processes occurring in the corresponding bulk. Accordingly, much attention has addressed interfacial chemistry to understand the physical origins engendering these differences, with factors such as dielectric constants[4–6], unique hydrogen-bonding networks[7,8], electric double layer formation[9–11], electric fields[12], etc., being invoked to rationalize the observed behavior.

The discovery that anions with weak hydration enthalpies can have strongly enhanced concentrations in the interfacial region is a relatively new phenomenon[13], which initially contradicted classical electrostatic theory and the interpretation of many "surface sensitive" experiments, e.g. the change in surface tension of water upon addition of salt[14]. Several inconsistencies have since been reconciled, and our understanding is now much more complete, with even the detailed mechanism of adsorption for weakly-hydrated, monovalent anions to water-hydrophobe interfaces being quite well established.[15–17] However, the picture remains unclear for strongly hydrated, polyatomic ions such as $CO_3^{2-}$ and $SO_4^{2-}$, which generally experience stronger image charge repulsion from the water-hydrophobe boundary. Given the importance of many strongly hydrated, polyoxy anions (e.g. $CO_3^{2-}$, $SO_4^{2-}$, $PO_4^{3-}$, $NO_3^-$, $XO_3^-$, X=Cl, Br, I and their respective acids) in atmospheric, environmental, and biological systems, further investigation of their behavior at the air-water interface is clearly warranted.

Our focus herein is on the centrally important carbonate system, which has been studied extensively, dating back to solubility experiments conducted over 100 years ago.[18] The hydration structure and dynamics of carbonate species have been characterized with a number of different methods, including MD simulation[19,20], quantum calculations[21–23], and X-ray spectroscopy[24–28]. Similarly, surface sensitive nonlinear spectroscopies have been used to study the behavior of these important anions. An early vibrational sum-frequency generation (vSFG) study by Tarbuck and Richmond reported that $Na_2CO_3$ perturbs the air-water interface more than does $NaHCO_3$; however, the authors did not comment on the relative surface affinities of the two anions.[3] Allen et al. conducted phase-sensitive measurements and found an increase in H-down oriented interfacial waters in the presence of carbonate and concluded that the sodium counter cation resides closer to the air-water interface than does the anion, and that bicarbonate resides closer to the surface than carbonate.[29] MD simulations supported these findings.[30,31] It has also been reported recently that the carbonate system does not form a well-ordered electric double layer.[32] We note that these vSFG measurements monitor solute-induced changes in the OH spectrum of water and are therefore *indirect* measures of the interfacial ion population.

Lam et al. recently reported the reversed fractionation of carbonate and bicarbonate anions at the air-water interface using Ambient Pressure Xray Photoelectron Spectroscopy (AP-XPS).[25] They found higher concentrations of doubly-charged carbonate ions than singly-charged bicarbonate in the near-interfacial region, contrasting previous surface-sensitive measurements. Without detailed theory to interpret this surprising behavior, and given that AP-XPS is not rigorously a surface-sensitive technique, we herein revisit the subject of the interfacial adsorption behavior of the carbonate system. In particular, we aim to clarify the discrepancies that exist in the literature regarding the relative surface affinity of the carbonate and bicarbonate anions at the air-water interface, and to propose a molecular-level picture for their observed surface behavior.

In this work, we employ resonantly enhanced deep-UV second harmonic generation (DUV-SHG) spectroscopy to *directly* probe the carbonate and bicarbonate anions at the air-water interface, with much higher surface specificity than the AP-XPS experiment. We find, contrary to classical electrostatic theory and in agreement with our previous AP-XPS measurements, that the doubly-charged carbonate anion does indeed exhibit a stronger preference for the air-water interface than does the singly-charged bicarbonate anion, and quantify this through the determination of the respective Gibbs free energies of adsorption. We performed molecular dynamics simulations (MD) to develop a detailed understanding of this interfacial behavior, and computed the population distribution and thermodynamic properties of $HCO_3^-$ and $CO_3^{2-}$ at the air-water interface. We find that transient agglomeration, in the case of $Na_2CO_3$ solutions, results in the formation of near-neutral, weakly solvated ion clusters which are highly surface active. Lastly, we simulate XPS binding energies of carbonate and bicarbonate at the air-water interface to help further interpret the experimental results from our previous AP-XPS experiments.

## SHG And the Langmuir Adsorption Model

To directly probe the carbonate and bicarbonate anions at the air-water interface, we use resonantly-enhanced deep-UV second harmonic generation (DUV-SHG) spectroscopy. SHG is a second-order nonlinear spectroscopic technique, wherein two photons at frequency ω combine to form one photon at frequency 2ω. Symmetry requirements under the electric dipole approximation dictate that a SH photon is only generated in a noncentrosymmetric environment (i.e an interface) and from a non-centrosymmetric molecule, rendering SHG a highly surface specific technique.[34] The SH intensity is governed by the equation:

$$I_{SH} \propto |\chi^{(2)}|^2 I_\omega^2 \qquad (1)$$

Here, $\chi^{(2)}$ is the second-order susceptibility, and $I_{SH}$ and $I_\omega$ are the intensities of the second harmonic and fundamental beams, respectively. Both the anions and water have their respective susceptibility tensors, and contribute to the overall signal:

$$I_{SH} \propto \left|\chi^{(2)}_{anion} + \chi^{(2)}_{water}\right|^2 I_\omega^2 \qquad (2)$$

The molecular responses for the anion and water are complex and contain both a real (non-resonant) and an imaginary (resonant) component. In the UV, the response from water is purely real. Therefore, under the two-photon-resonant conditions employed here, the resonant term from the anion should dominate the overall measured SH intensity. In the low solute concentration regime and/or under non-resonant conditions, the signal from water and the solute need to be considered equally, which will be discussed in more detail later. We can express the resonant term for the anion as:

$$\chi^{(2)}_{anion} \propto N_{anion} <\beta_{anion}> \quad (3)$$

Here $\beta_{anion}$ is the orientationally averaged hyperpolarizability and $N$ represents the number density of the anion at the interface. Equation 3 emphasizes that changes in the number density, as well as the orientation of the resonant species will change the measured SH signal.

Because the signal depends on the relative concentration of the adsorbing solute, we can apply a simple Langmuir model to fit our experimental data. The use of a Langmuir model for SHG is well established[33] and only the important features are highlighted here. For a more thorough derivation, we refer the reader to the supporting information. We parameterize the second order susceptibility as:

$$|\chi^{(2)}|^2 \propto |A + N_s(B + iC)|^2 \quad (4)$$

Here, A represents the non-resonant response from water, $N_s$ is the number density of adsorbed anion, and $B$ and $C$ are the non-resonant and resonant susceptibilities of the anion, respectively. $N_s$ is determined from a kinetic description of surface exchange, and related to the bulk anion concentration, $X_{SCN^-}$, and Gibbs free energy of adsorption, $\Delta G$:

$$\frac{I_{2\omega}}{I_\omega^2} = |\chi^{(2)}|^2 \propto \left(A + B\frac{X_{SCN^-}}{(1-X_{SCN^-})e^{\frac{\Delta G}{RT}} + X_{SCN^-}}\right)^2 + \left(C\frac{X_{SCN^-}}{(1-X_{SCN^-})e^{\frac{\Delta G}{RT}} + X_{SCN^-}}\right)^2 \quad (5)$$

Thus, by measuring the second harmonic intensity as a function of anion concentration, the Gibbs free energy of adsorption can be determined.

## Results / Discussion

### Second-Harmonic Generation Spectroscopy

Figure 1a shows the bulk absorption spectra of Na$_2$CO$_3$ and NaHCO$_3$ in water at 293 K, with both anions exhibiting molecular $\pi\pi^*$ transitions in the deep UV. The geometries of the carbonate and bicarbonate anions in solution have been well characterized; quantum chemical calculations[21] and neutron diffraction[33] showed that the normally planar $CO_3^{2-}$ anion exhibits broken symmetry in aqueous solution (D$_{3h}$ →

C$_{3v}$) due to interactions with the solvent. Raman spectroscopy studies in solution have also concluded that the first shell solvation environment is indeed itself asymmetric.[34] These properties of the carbonate and bicarbonate anions allow us to directly probe their relative concentrations in the interfacial region. A cartoon depiction of the experimental design is shown in Figure 1b.

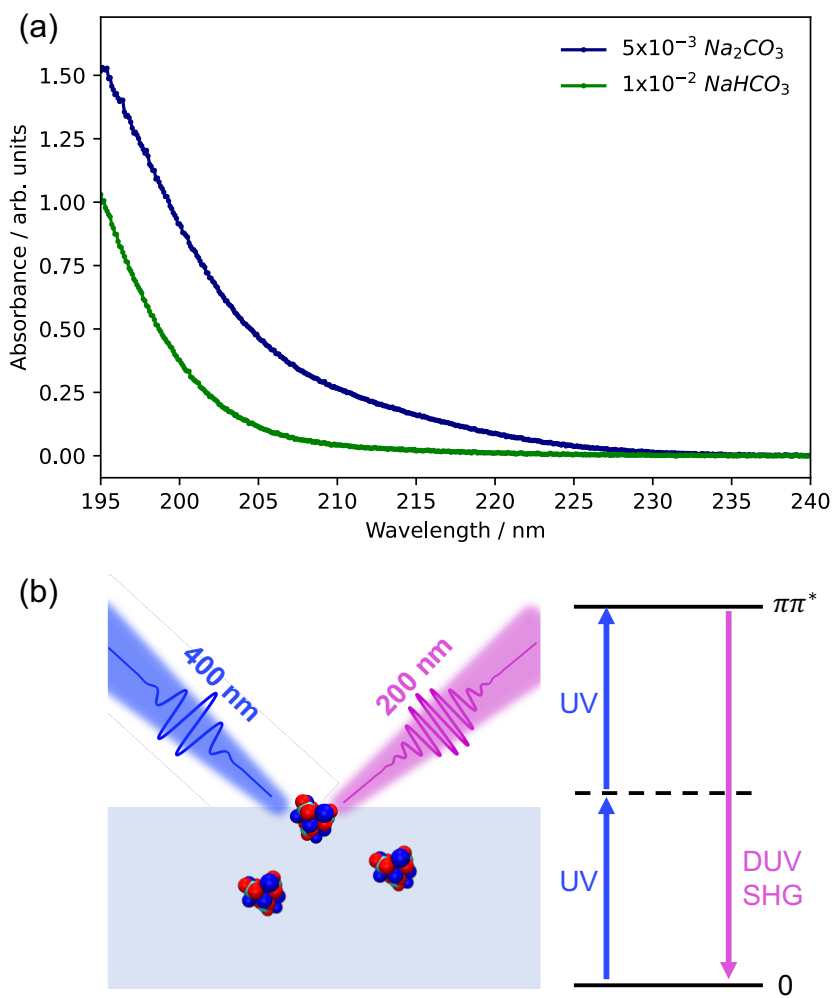

**Figure 1: (a)** Linear absorption spectra for 5x10$^{-3}$M Na$_2$CO$_3$ (blue line) and 1x10$^{-2}$M NaHCO$_3$ (green line) measured at 293 K. **(b)** Cartoon depiction of the experimental design showing generation of a second harmonic photon from a carbonate cluster residing at the liquid water surface. The energy level diagram highlights the resonant enhancement from being two-photon resonant with the molecular $\pi\pi^*$ transition of the carbonate and bicarbonate anions.

Figure 2a shows the normalized (relative to pure water) SH response of the carbonate anion at the air-water interface for 0-0.036 mole fraction (0-2.0 M) solutions of sodium carbonate. The intensities are measured in two polarization combinations (s-in, p-out and p-in, p-out), with both polarizations giving a fairly weak second-order response, ca. twice the response of pure water at a concentration of 2.0 M bulk concentration; this is likely due to the small extinction coefficient at 200 nm for carbonate when compared to other surface active anions such as $SCN^-$, which exhibits a much stronger SH response.[35] In the low concentration region from 0 – 0.0045 mole fraction (0-0.175 M), the measured SH response is slightly below or equal to that of pure water, indicating no appreciable amount of carbonate in the probe depth of SHG, as well as minimal re-orientation of interfacial water molecules.

Electrolyte solutions with normalized SHG values below the response of pure water have been observed previously in the low concentration limit, and the mechanism for this response is still debated in the literature. Explanations such as destructive interference between the resonant signal from the electrolyte and the non-resonant signal from water have been implicated[36,37]; however, more recent studies seem to point towards long range correlations extending from bulk water, induced by the electrolyte, as the cause of this behavior[38,39]. This phenomenon is most relevant for the discussion of the Jones-Ray Effect in sub-molar concentrations, and is not a focus of the present work.

Between 0.0045 – 0.018 mole fraction (0.175 – 1.0 M), the SH response begins to increase linearly with anion concentration, owing to resonant enhancement of carbonate as it accumulates in the interface, as well as alignment of interfacial water from the electric field of the anion. The latter contribution effectively increases the hyperpolarizability of water (see Eqn. 3), and is akin to a "thickening" of the interface.[40] Above 0.018 mole fraction (1.0-2.0 M), the SH response becomes asymptotic, as the surface becomes "saturated" and cannot accommodate any additional carbonate anions. The concentration range used in these experiments is limited by the solubility of sodium carbonate, which at room temperature is ~2M. The Langmuir adsorption model, as employed herein, gives an unconstrained, best fit to the data with a Gibbs free energy of adsorption of -11.1 ± 0.7 kJ/mol for both p-in, p-out and s-in, p-out polarization, with the uncertainty being one standard deviation.

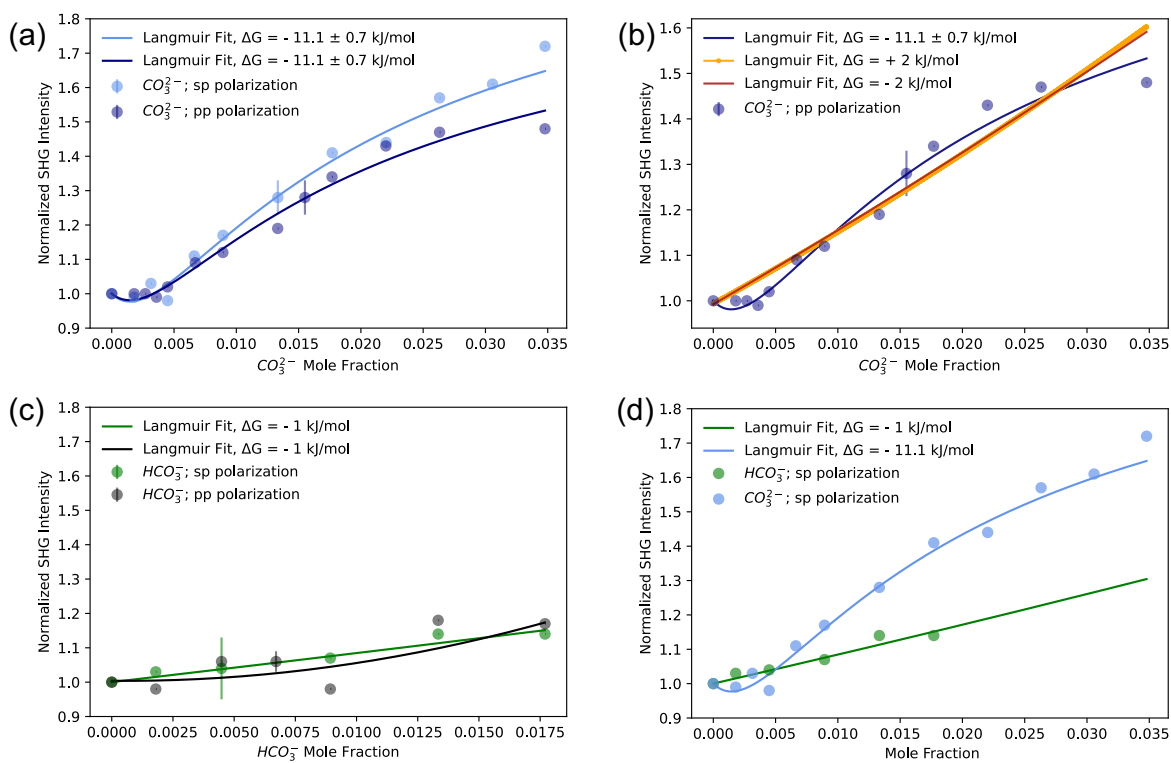

**Figure 2**: **(a)** Normalized SHG intensities for the carbonate anion at the air-water interface, measured at a SH wavelength of 200nm for *s*-in, *p*-out (light blue circles) and *p*-in, *p*-out (dark blue circles). For clarity, only one representative error bar for each data set is shown, which corresponds to one standard deviation. Each data set is fit to a Langmuir adsorption model, yielding Gibb's free energies of adsorption of -11.1 ± 0.7 kJ/mol. **(b)** Langmuir adsorption model shown for three different adsorption regimes for the carbonate anion: strong adsorption (dark blue), weak adsorption (red line), and repulsion (orange line). **(c)** Normalized SHG intensities of the bicarbonate anion at the air-water interface, measured at a SH wavelength of 200nm for *s*-in, *p*-out (green circles) and *p*-in, *p*-out (black circles). A Langmuir fit with the Gibb's free energy constrained to -1kJ/mol is shown for each data set. This value represents a minimum threshold for the Gibb's free energy. **(d)** SHG intensities for carbonate and bicarbonate. Extrapolation of the Gibbs free energy for the bicarbonate fit to concentrations of 2M reveals as SH response significantly weaker than that for carbonate.

In Figure 2b right panel, we compare the best fit of the Langmuir model for the carbonate anion (-11.1 ± 0.7 kJ/mol) with the output of the Langmuir model in the weak adsorption regime ( -2 kJ/mol, orange line) and the repulsive regime (+2 kJ/mol, blue line). In order to obtain these fits corresponding to weak and repulsive adsorption, the Gibbs free energy was constrained to these values and the fitting parameters A,B,C were simultaneously solved for. Both the -2 and +2 kJ/mol fits increase linearly with increasing bulk anion concentration, and clearly do not adequately represent the observed curvature in the experimental data. In the weak and repulsive adsorption regime, the non-resonant parameter B increases, and the resonant parameter, C, becomes very small or even negative (see SI Table 1). Under the experimental conditions employed here, wherein the SH wavelength is resonant with the anion, these fitting parameters are unphysical - implying that the Langmuir model does not produce a representative fit to the data.

Figure 2c shows the normalized SH response of the bicarbonate anion from 0 – 0.0175 mole fraction (0 – 1.0 M) solutions of sodium bicarbonate. With both s- and p-input polarization, there is a weak, linear increase in the SH signal with increasing bulk anion concentration. This behavior, as mentioned above, is ascribed to a thickening of the interfacial layer via alignment of surface waters, and has been observed before in non-resonant studies of anions that are known to be repelled from the interface, e.g. the fluoride anion, $F^-$.[40]

A fit of the bicarbonate data to the Langmuir model does not converge, and therefore the Gibbs free energy cannot be determined from an unconstrained fit. Instead, in Figure 2c, constrained fits with Gibbs free energies of -1 kJ/mol are shown for both data sets. This value of the Gibbs free energy corresponds to a weak adsorption event. More negative values of the Gibbs free energy (i.e. stronger adsorption) induce curvature in the fit that is not present in the experimental data; therefore, an upper limit of -1 kJ/mol is established. Less negative, and positive, values of the Gibb's free energy produce results that are identical to those of the -1 kJ/mol fit, viz. a shallow linear increase with increasing anion concentration. Therefore, while we cannot quantitatively determine the Gibbs free energy of adsorption for the bicarbonate anion, we can establish a range of values, with an upper limit of -1 kJ / mol, and less favorable values of adsorption being more likely.

Our DUV-SHG measurements indicate that the carbonate anion has a much stronger surface affinity for the air-water interface than does the bicarbonate anion, as shown in Figure 2d. On the basis of hydration free energy ( $\Delta G_{hyd}(CO_3^{2-}) = -1315 \frac{kJ}{mol}$ and $\Delta G_{hyd}(HCO_3^-) = -335 \frac{kJ}{mol}$ )[41], this behavior is indeed puzzling. It has been shown by both experiment and theory, that weakly-hydrated anions tend to be surface active, whereas strongly hydrated anions tend to be surface repelled, roughly following the Hofmeister series.[42,43] For example, anions that have been shown to have enhanced concentrations in the interfacial region, e.g. $I^-$ and $SCN^-$, have hydration free energies of ca. -280 kJ/mol. The surface adsorption of such a strongly-hydrated anion as carbonate indeed seems unlikely, and we turn to MD simulations of these systems to help interpret these surprising experimental results.

## Molecular Dynamics Simulations

Our analysis of molecular dynamics trajectories shows significant differences in the solvation structure between $HCO_3^-$ and $CO_3^{2-}$ species. Specifically, we find for solutions of $Na_2CO_3$ that $CO_3^{2-}$ tends to form agglomerates with $Na^+$, resulting in clusters or chains, while in solutions of $NaHCO_3$, the $HCO_3^-$ anion remains isolated (Figure 3a and 3b). In Figures 3c and 3d, we plot the coordination number of different atoms in carbonate and bicarbonate solutions as a function of radial distance. The $CO_3^{2-}$ molecules on average have much higher coordination number with $Na^+$ and neighboring carbonates than does $HCO_3^-$. Additionally, Figure 3d highlights that the coordination numbers on average for both $CO_3^{2-}$ and $HCO_3^-$ with water are equal at short distances (up to 2 Å), and then larger for $HCO_3^-$ at longer distances. Given the large difference in hydration free energy for these anions, one would expect the carbonate-water coordination number to be significantly higher than for bicarbonate. Indeed, this has been observed in the radial distribution functions for carbonate and bicarbonate anions from MD simulation, with no counter cations present in the solutions.[20] The observed difference results from increased electron density on the oxygens of carbonate, coupled with higher prevalence of sodium cations, which compete with water for interaction in the solvation environment. In the case of bicarbonate, which has reduced electron density and half the number of sodium cations, interaction with the water is more likely. This behavior can be explained through the formation of a weakly solvated agglomerate for $CO_3^{2-}$. The $Na^+|CO_3^{2-}$ agglomerate, which is relatively neutral and weakly hydrated, leads to a higher propensity for the interface relative to the charged $HCO_3^-$.

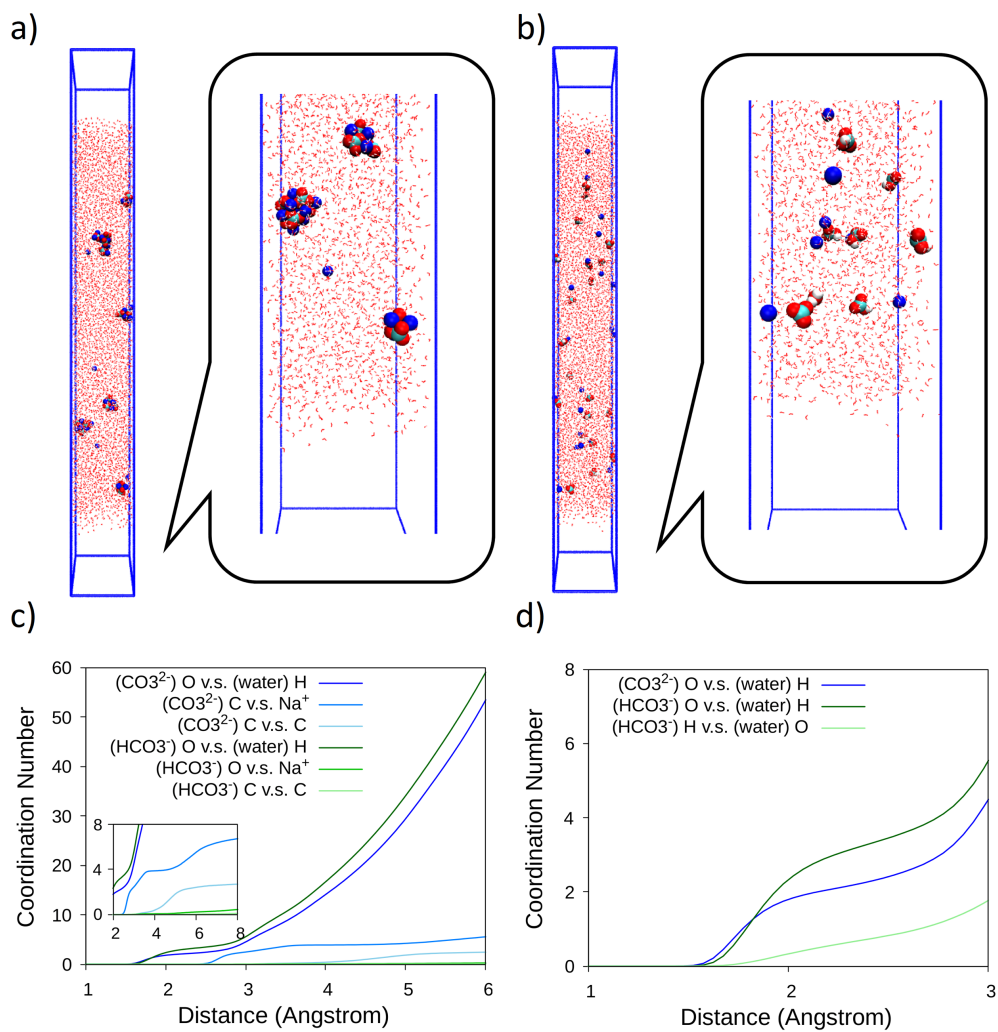

**Figure 3**: Molecular dynamics simulation snapshots for solutions of **(a)** $Na_2CO_3$ and **(b)** $NaHCO_3$. Panel **(a)** highlights the formation of agglomerates in solution. **(c)** and **(d)** show the coordination number for different atoms in carbonate and bicarbonate solution as a function of radial distance.

That this agglomerate shows higher surface propensity than its non-clustered, charged counterpart, as well as bicarbonate, is in agreement with current descriptions of ion adsorption. Saykally and Geissler have shown that there is an enthalpic benefit for weakly-solvated ions to partition to the air-water interface, viz the shedding of one or two solvating waters which can then form stronger hydrogen bonds as part of the bulk water network. How capillary waves are influenced in the presence of this large agglomerate is still an open question, and could have important consequences for processes such as gas adsorption onto the liquid surface and evaporation kinetics.

Further insights into the surface enhancement of $CO_3^{2-}$ were provided by means of 2D accelerated dynamics simulations. We consider a 2D free energy surface defined by the distance of the center of mass of a carbonate ion to the interface (Z = 0) and the C – C coordination number of the carbonate species given by Equation (6)

$$\sum_i \sum_{j \neq i} \frac{1 - \left(\frac{r_{ij} - d_0}{r_0}\right)^n}{1 - \left(\frac{r_{ij} - d_0}{r_0}\right)^m} \qquad (6)$$

The 2D free energy diagrams are shown in Figure 4. The energy requirement for fully solvated, isolated carbonate species to move from the bulk region to the air-water interface is lower in the case of $CO_3^{2-}$. This is due to the thermodynamically favorable formation of agglomerates, which leads to a lower energy barrier. In addition, their final agglomerated state is more stable than the $HCO_3^-$ counterpart.

We find a propensity for the carbonate species to agglomerate near the interface, forming near-neutral 2 – 3 $Na^+/CO_3^{2-}$ clusters. These clusters exist in a steady-state equilibrium with more extended, polymer-like chains. The chain state is driven by the binding of water molecules but is less favored at room temperature. Near the interface, the agglomerates incur a reduced H-bonding penalty, since interfacial waters are less tightly bound and have higher entropy, facilitating the carbonate agglomerates existence at the interface. On the other hand, due to the singly-charged bicarbonate's reduced electrostatic binding, it remains isolated and thus repelled from the surface. Therefore, we establish that the local nanostructure of the carbonate anion plays a crucial role in the final solvation state at the interface.

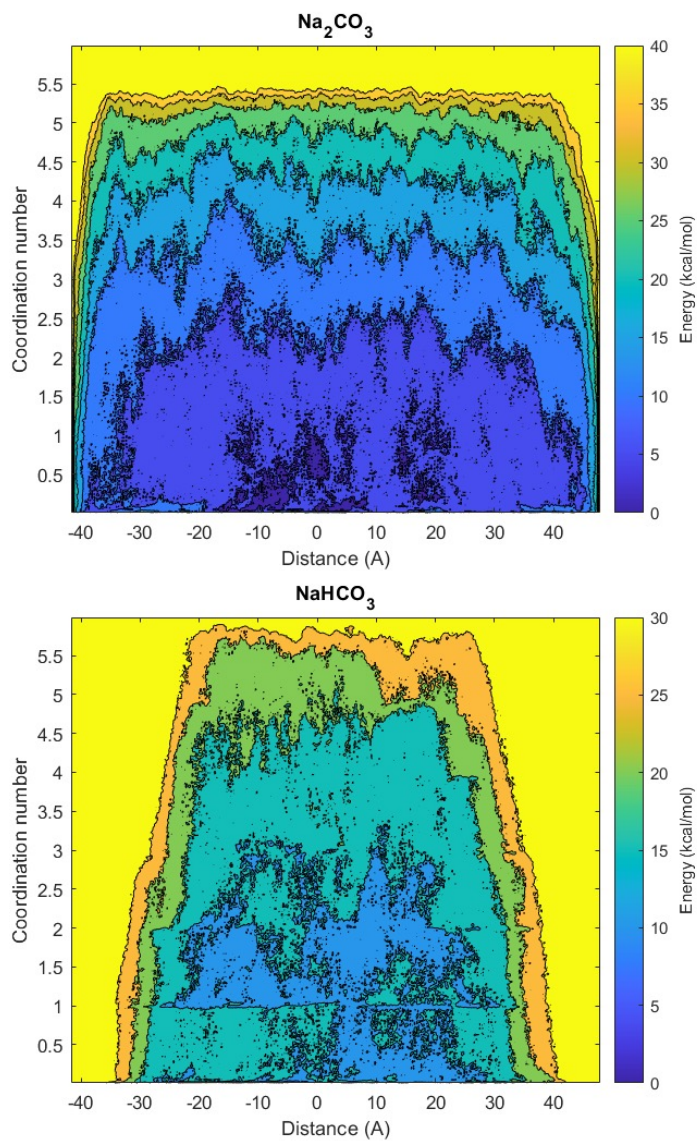

**Figure 4:** 2D Free energy surface of carbonate species as a function of distance from the bulk and coordination number. The coordination number is defined as the number of carbon atoms with in a distance of 3.5 Å.

### Confirmation of AP-XPS Measurements

As mentioned earlier, Lam et al. used AP-XPS to directly probe the carbon K-edge of the carbonate and bicarbonate anions at the air-water interface and measure their relative concentrations (Figure 5b).[25] By tuning the input photon energy, they exploited different attenuation lengths of the emitted photoelectron and were able to achieve depth profiling as shallow as ~2 nm. Their findings agreed with those of the present study, viz. that $CO_3^{2-}$ is more prevalent in the interface than $HCO_3^-$. However, as was noted above, AP-XPS probes deeper into the interface than do second-order spectroscopies. Even with a minimum electron attenuation length of 2 nm, there is significant signal measured from probe depths as deep as 5 nm.

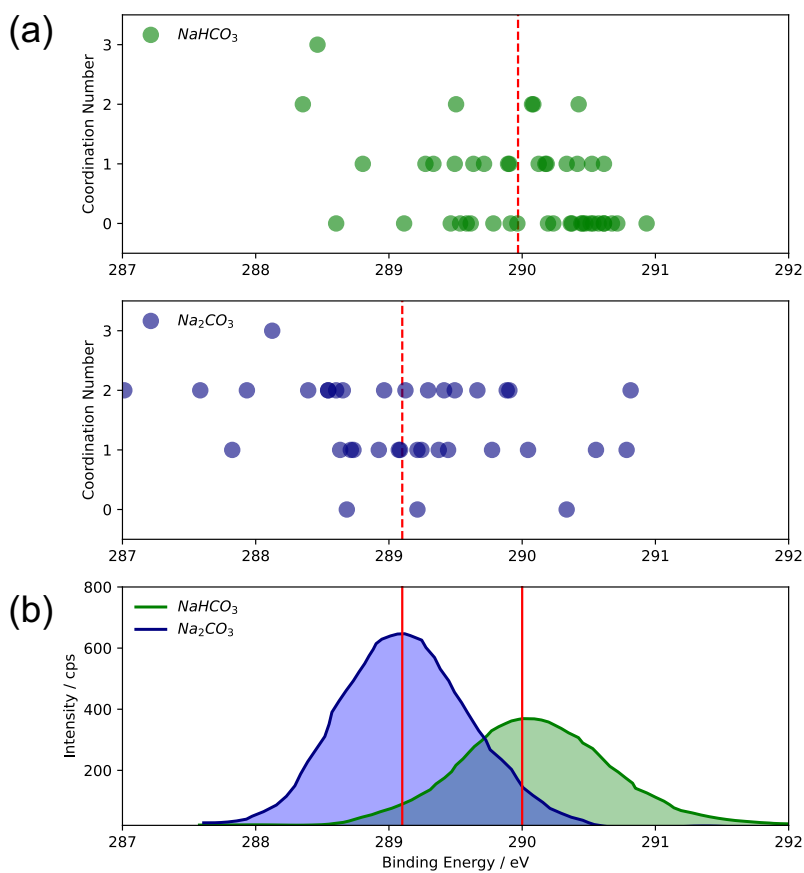

**Figure 5:** **(a)** Simulated XPS binding energies for C(1s) excitation of carbonate and bicarbonate at the air-water interface. Individual data points indicate the calculated binding energy and the associated coordination number to the anion. Calculated binding energies are "energy-aligned" to the experimental carbonate peak at 289.1 eV. **(b)** Xray Photoemission C(1s) binding energies with an incident photon energy of 490 eV, from 0.5 M solutions of $NaHCO_3$ and $Na_2CO_3$. Data in the bottom panel is reproduced from Lam et al. with permission from the authors.

An important consideration in XPS is the local environment of the excited atom, which has a significant effect on the measured binding energy. Given the different solvation environments between the bulk and surface, one might expect surface effects to modulate the spectra, viz binding energy shifts or transition narrowing/broadening. These shifts could influence the spectral fitting and thereby the relative concentrations of carbonate/bicarbonate reported by Lam et al. In order to address this concern, we calculate the C K-edge binding energies for carbonate and bicarbonate at the air-water interface, shown in Figure 5a. Our calculations reveal that the XPS spectra were invariant to displacement of the anion relative to the air-water interface, as well as to the coordination number of the carbonate species. In Figure 5b (bottom panel) the AP-XPS spectra from Lam et al are reproduced. The relative shift between the experimental carbonate/bicarbonate spectral signatures agrees well with the shift in the calculated binding energies, with a difference of only ~0.1 eV, indicating indeed that the spectral fitting and relative concentrations reported by Lam are of the distinct carbonate/bicarbonate peaks, and not some convolution of the two due to spectral shifts. These findings imply that the results of the experiment are mainly dictated by the thermodynamics properties and concentration of carbonate species at the interface.

**Implications for Atmospheric, Environmental, and Biological Sciences:**

The relative partitioning of the carbonate and bicarbonate ions to the air-water interface is of significant environmental and atmospheric importance, for example in atmospheric aerosol droplet and ocean-surface chemistry. A key consideration to this subject, which has not been discussed in the present work, is the adsorption and dissolution of atmospheric $CO_2$ from the gas phase. We have previously addressed the solvation and hydrolysis of the species involved in this process by X-ray absorption spectroscopy[24,26,27], and molecular simulations have highlighted the importance that the interface plays in the adsorption of a polyoxy anion to liquid surfaces.[44]

The ocean serves as a large sink of $CO_2$ from the atmosphere, with approximately 30% of all anthropogenic $CO_2$ emissions being absorbed[45,46].

Following solvation into the liquid, carbon dioxide gas hydrolyzes to form carbonic acid, which can further exchange protons to form bicarbonate and carbonate, which is a pH dependent process[47]:

$$H_2CO_3(aq) + H_2O \leftrightarrow HCO_3^-(aq) + H_3O^+(aq) \quad (pK_a = 6.4) \quad (7)$$
$$HCO_3^-(aq) + H_2O \leftrightarrow CO_3^{2-}(aq) + H_3O^+(aq) \quad (pK_a = 10.3) \quad (8)$$

Thus, an increase in atmospheric $CO_2$ could greatly impact marine ecosystems via ocean acidification, as given by Eqn. 7 and 8 [48,49], and will shift the ratio of carbonate and bicarbonate anions present in ocean waters and at the air-ocean interface.

While the ocean has a large degree of buffering capacity, this effect could be quite drastic inside of aerosol droplets, which have high surface area to volume ratios. We have shown here that carbonate resides at the interface in higher quantity than does bicarbonate, and expect this partitioning to play an important role in influencing aerosol chemistry. It has been postulated that reaction kinetics can be enhanced at the surfaces of aerosol droplets[50,51], hence, the change in pH inside of droplets (and subsequent change in surface composition/structure) could disrupt or potentially further enhance these fact reactions. Furthermore, since mammalian respiration systems are also dependent on the buffering capacity of the carbonate system and help to regulate changes in blood pH, a rigorous understanding of carbonate and bicarbonate affinities for, and reactions at, the air-water interface would further illuminate these important topics.

## Conclusions

In this work, we employed resonantly-enhanced DUV-SHG to directly probe the carbonate and bicarbonate anions at the air-water interface. By fitting the concentration dependence with a Langmuir adsorption model, we determined that the doubly-charged carbonate anion adsorbs more strongly to the surface than does singly-charged bicarbonate, in conflict with widely used models and general expectations. These measurements support our previous study of the carbonate system conducted by AP-XPS. We also describe MD simulations, which revealed that agglomerate formation of the highly charged $CO_3^{2-}$ with $Na^+$ counterions is the driving force for its surface propensity, and that this same behavior is not found for the singly charged $HCO_3^-$. We hope that these new results will inspire further experiments and modeling for deeper insight into this vitally important chemical system.

# Supporting Information: Agglomeration Drives The Reversed Fractionation of Aqueous Carbonate and Bicarbonate at the Air-Water Interface

**Experimental**

*Materials:* All glassware was soaked overnight in Alnochromix (Alconox Inc.) cleaning solution and rinsed thoroughly with ultrapure (18.2 MΩ) water. Solutions were prepared by dissolving reagent grade sodium carbonate (Sigma Aldrich, >99% purity) and sodium bicarbonate (Sigma Aldrich, >99% purity) in ultrapure water.

*Second Harmonic Generation Spectroscopy:* The experimental design has been described in detail elsewhere[1], and only a brief description is given here. The output from a Ti-S regenerative amplifier (Spectra Physics Spitfire, 4 mJ, 100fs, 1kHz) is directed through a BBO crystal to generate light at 400 nm. The 400 nm light is then focused onto the surface of the solution using a $f = 100$mm lens at an angle of 60° relative to the surface normal. The fundamental and SH light are collected with a collimating lens, and the fundamental light is spectrally filtered using a laser line mirror and Pellin-Broca prism. The SH photons are then detected using a gated boxcar integrator (Stanford Research Systems), a monochromator (Acton SpectraPro 2150i), and a photomultiplier (Hamamatsu, R7154PHA) for photon counting. All SHG measurements are normalized relative to the SH response of pure water. Input polarization was controlled using a half-wave plate and a polarizer.

*Derivation of Langmuir Adsorption Model for Use in SHG*

The second harmonic (SH) intensity can be written as:

$$I_{SH} \propto |\chi^{(2)}|^2 I_\omega^2 \quad (1)$$

Here $\chi^{(2)}$ is the second-order susceptibility and $I_\omega$ is the intensity of the fundamental light. $\chi^{(2)}$ is complex valued, and has a contribution from each component in the system:

$$\frac{I_{SH}}{I_\omega^2} \propto \left|\chi^{(2)}_{water} + \chi^{(2)}_{anion}\right|^2 \quad (2)$$

$\chi^{(2)}$ is dependent on the number density, $N$, as well as the effective hyperpolarizability, $\beta^{eff}$

$$\frac{I_{SH}}{I_\omega^2} \propto \left|N_{water} * \beta^{eff}_{water} + N_{anion} * \beta^{eff}_{anion}\right|^2 \quad (3)$$

Water does not give a strong SH response, and we assume its contribution to the signal is real. However, because carbonate and bicarbonate are resonant at the SH wavelength they have both a real and imaginary component. We can group these contributions accordingly

$$\frac{I_{SH}}{I_\omega^2} \propto \left(N_{water} * \beta^{eff}_{water} + N_{anion} * Re\{\beta^{eff}_{anion}\}\right)^2 + \left(N_{anion} * Im\{\beta^{eff}_{anion}\}\right)^2 \quad (4)$$

and switch to using concentrations by dividing each term by $N_{water}$, since $N_{anion}$ / $N_{water}$ is a concentration:

$$\frac{I_{SH}}{I_\omega^2} \propto \left(\beta^{eff}_{water} + \frac{N_{anion}}{N_{water}} * Re\{\beta^{eff}_{anion}\}\right)^2 + \left(\frac{N_{anion}}{N_{water}} * Im\{\beta^{eff}_{anion}\}\right)^2 \quad (5)$$

We simplify the expression to:

$$\frac{I_{SH}}{I_\omega^2} \propto (A + N_s * B)^2 + (N_s * C)^2 \quad (6)$$

Here, the weak system response from water is represented by the non-resonant term A, B is the real component of the anion susceptibility, C is the imaginary component of the anion susceptibility, and $N_s$ is the concentration of surface active anions.

To develop an expression for the number of surface active anions, $N_s$, we use the Langmuir adsorption model, where an anion in the bulk can exchange with a water at the surface:

$$Anion_{bulk} + water_{surf} \leftrightarrow Anion_{surf} + water_{bulk} \quad (7)$$

Here, the subscripts *surf* and *bulk* refer to an anion/water molecule occupying a surface site or bulk site, respectively. We can write the equilibrium expression according to the concentration of each species as:

$$K_{ads} = \frac{[Anion]_{surf} * [water]_{bulk}}{[Anion]_{bulk} * [water]_{surf}} \quad (8)$$

If we assume a maximum number of surface sites, [sites]$_{max}$ the expression becomes:

$$K_{ads} = \frac{[Anion]_{surf} * [water]_{bulk}}{[Anion]_{bulk} * ([sites]_{max} - [Anion]_{surf})} \quad (9)$$

Rearranging for $[Anion]_{surf}$ gives:

$$N_s = [Anion]_{surf} = [sites]_{max} * \frac{[Anion]_{bulk}}{[water]_{bulk} * K_{ads}^{-1} + [Anion]_{bulk}} \quad (10)$$

Substituting $N_s$ back into Equation (6), changing to mole fraction and using the relationship between the equilibrium adsorption rate and the Gibbs free energy gives

$$\frac{I_{2\omega}}{I_\omega^2} \propto |\chi^{(2)}|^2 \propto \left( A + B \frac{X_{anion}}{(1 - X_{anion})e^{\frac{\Delta G}{RT}} + X_{anion}} \right)^2 + \left( C \frac{X_{anion}}{(1 - X_{anion})e^{\frac{\Delta G}{RT}} + X_{anion}} \right)^2 \quad (11)$$

which relates the observed SH intensity to the bulk anion concentration ($X_{anion}$) and the Gibbs free energy of adsorption ($\Delta G$).

SI Table 1: Fitting Parameters from Langmuir Model of Adsorption for the carbonate anion.

| Anion | $\Delta G \left(\frac{kJ}{mol}\right)$ | A | B | C |
|---|---|---|---|---|
| $CO_3^{2-}(pp)$ | -11.1 ± 0.7 | 1.00 | 0.164 | 1.12 |
| $CO_3^{2-}(sp)$ | -11.1 ± 0.7 | 1.00 | 0.19 | 1.23 |
| $CO_3^{2-}(pp)$ | -2 ** | 0.99 | 3.5 | -3.5x10$^{-5}$ |
| $CO_3^{2-}(pp)$ | +2** | 0.99 | 17.2 | 4.2x10$^{-4}$ |

** indicates that the Gibb's free energy of adsorption was constrained during the fit, and parameters A,B,C were simultaneously solved for.

## Simulation

### MD simulation

The Large-scale Atomic/Molecular Massively Parallel Simulator (LAMMPS)[2] was adopted in this work to perform molecular dynamics simulations (MD). The carbonate ($CO_3^{2-}$) atomic charge, van der Waals interactions, and intra-molecular motions were described by Wang et al.'s work.[3] The bicarbonate ($HCO_3^-$) van der Waals interaction adopted Wang et al.'s results[3] as well, combined with zero epsilon value for the hydrogen atoms. The $HCO_3^-$ atomic charge and intra-molecular motions were taken from Demichelis et al.'s work.[4] Water molecules were based on the TIP3P forcefield[5] with SHAKE algorithm to provide rigid bond and rigid angle, and sodium ions ($Na^+$) were taken from Joung et al.'s work. [6]

A simulated cell with 30Å x 30Å x 260Å in the x, y, z dimensions was set with periodic conditions in the x- and y-boundaries, and finite condition in the z-boundary. The concentration of carbonate species were set to 0.2 M which corresponds to 20 molecules of Na₂CO₃/NaHCO₃ and 5800 water molecules. In the z-direction, there was ~40Å vacuum space below and above the $CO_3^{2-}/HCO_3^-|Na^+|$ aqueous solution to establish the water-vapor interface. Each MD model was initiated with 500 steps CG minimization, followed by 10 ps Nose-Hoover thermostat (NVT) to heat the system to 298K. Afterwards, 5 ns NVT was applied to well produce molecule dynamics trajectories. To avoid the slab-slab interactions, a volfactor value of 2.0 was used. In addition, Lennard-Jones 9-3 potential walls with ε=σ=1.0 and a cutoff of 2.5 Å were added at the z-boundary edges. A particle-particle particle-mesh solver was applied with 10⁻⁴ tolerance for long-range interactions. The timestep was 1.0 fs and the temperature dump factor was 100.

### Solvation energy of $CO_3^{2-}$, $HCO_3^-$ and $Na^+$

We employed a Free Energy Perturbation model (FEP) to determine the solvation free energy of ions in aqueous solution. A coupling parameter, λ (0 ≤ λ ≤ 1), is introduced into the simulation. The coupling parameter controls interactions between water and the solvated species. By slowly varying this coupling parameter, the solvation energy can be obtained from equation below.

$$\Delta A_0^1 = \sum_{i=0}^{n-1} \Delta_{\lambda_i}^{\lambda_{i+1}} A = -kT \sum_{i=0}^{n-1} ln \langle exp\left(-\frac{U(\lambda_{i+1})-U(\lambda_i)}{kT}\right)\rangle_{\lambda_i}$$

Here, A is the solvation energy and U is the free energy of the system.

### Accelerated Meta-Dynamics Simulation

We explored the rugged free energy landscape for carbonate systems using well-tempered[7] multiple walker[8] metadynamics.[8–10] In all cases, we use 35 walkers, each initiated from various points along the equilibrium production MD trajectory. In each simulation, MW-wt-MetaD biases were constructed as follows: Gaussian functions were deposited every 0.5 ps with an initial height of 277/T x 1.0 kcal/mol. The bias factor [γ = (T + ΔT) / T] was set to 5. We monitored convergence by calculating the free energy profiles every 1ns and found that ~ 30ns was reasonable in most cases. All simulations were performed using LAMMPS and Plumed 2.5.[11–13,14]

### XPS binding energy simulation

Due to the relatively large size of the system, we performed another set of MD simulation on a smaller system of 15Å x 15Å x 60Å to generate snapshots for electronic structure calculations. Each representative structure has the targeted carbon containing chemical species at the air-water interface and in the bulk. We employed ARES package to simulate XPS binding energies of C K-edge excitation in carbonate aqueous solution.[15] ARES is a realspace KS-DFT package utilizing Chebyshev subspace filtering, finite difference real-space grid and pseudopotential scheme, advanced KS-DFT solvers, versatile non-periodic/periodic boundary conditions, and highly parallelizable features making the simulation of our system possible. A total of 8 snapshots at the air/water interface from molecular dynamics simulation were taken as an input structure for the DFT simulation. The binding energy expression is obtained by SCF within KS-DFT comparing between the initial ground state and final excited state. The XPS binding energy can be expressed below:

$$E_b = E^{N-1}[n_F] - E^N[n_I], \qquad (1)$$

where $E^{N-1}[n_F]$ and $E^N[n_I]$ are the total energy functionals of the final and initial electron density, respectively. Specifically, two PPs for carbon atoms with pseudo core and core-hole respectively are generated using the FHI98PP code[16] for the initial and final state calculations, respectively. Under the fully-screened core-hole assumption, only the traditional self-consistent iterations are required for both initial and final state calculations. All calculations were performed using PBE functionals. Grid spacing of 0.15 Å was adopted to achieve well-converged total energies (1meV/atom). The total energy convergence criterion of self-consistent field iteration is within 0.1 meV.